\documentclass[12pt]{article}
\usepackage[dvips]{color}
\usepackage{epsfig}
\usepackage{amsmath}
\usepackage{cite}
\usepackage{graphicx}
\usepackage{color}
\usepackage{subfigure}
\def\Box{\hbox{$\rlap{$\sqcup$}\sqcap$}}
\def\Box{\hbox{$\rlap{$\sqcup$}\sqcap$}}
\begin{document}

\begin{center}
\Large{\bf {Swampland Conjectures in\\ Hybrid metric-Palatini gravity}}\\
\small \vspace{1cm} {\bf J. Sadeghi$^{\star}$\footnote {Email:~~~pouriya@ipm.ir}}, \quad
{\bf S. Noori Gashti$^{\star}$\footnote {Email:~~~saeed.noorigashti@stu.umz.ac.ir}}, \quad
{\bf F. Darabi$^{\dagger}$\footnote {Email:~~~f.darabi@azaruniv.ac.ir}},
\quad
\\
\vspace{0.5cm}$^{\star}${Department of Physics, Faculty of Basic
Sciences,\\
University of Mazandaran
P. O. Box 47416-95447, Babolsar, Iran}\\
$^{\dagger}${Department of Physics, Faculty of Basic
Sciences,\\
Azarbaijan Shahid Madani University
P. O. Box 53714-161, Tabriz, Iran}
\small \vspace{1cm}
\end{center}
\begin{abstract}
In this paper, we study a hybrid combination of Einstein-Hilbert action with curvature scalar $R$, and a function $f(\mathcal{R})$ in Palatini gravity within the context of inflationary scenario, from the Swampland conjecture point of view. This hybrid model has been paid
attention in recent cosmological studies, and its applications have been widely studied in the literature. In this regard, using the Swampland conjecture ({{using ($ C_ {1} $) as the first component of dS swampland conjecture, which is obtained from the first derivative of the potential upon the potential and ($ C_ {2} $) as the second component which is acquired from the second derivative of the potential upon the potential}}), we investigate  the cosmological implications of the present gravity theory, with
a suitable potential, in
the framework of inflationary scenario to obtain cosmological quantities such as slow-roll parameter, scalar spectral index $(n_{s})$, tensor-to-scalar ratio ($r_{s}$), and then compare them with the cosmological observations. Moreover, we compare the compatibility or incompatibility of the model with observable data, such as Planck, by applying Swampland conjecture  to $r_{s}-n_{s}$ , $C_{1,2}-n_{s}$ and $C_{1,2}-r_{s}$ plots.\\\\
Keywords: Einstein-Hilbert action, Palatini gravity, Swampland Conjecture.
\end{abstract}
\newpage
\section{Introduction}

One of the most exciting issues is understanding the universe and the formation of its structures\cite{eee,fff,ggg}. The recent observations on the accelerated expansion of the universe provided great interests in correcting the geometry sector of Einstein's field equations.\cite{1,2,3} Many modified theories from this perspective were unveiled, and a lot of researches on effective $f(R)$ theories and their different versions was accomplished\cite{4,5,6,7,8,9,10,11}. It has recently been shown that the metric formalism of these theories is different from its  Palatini formalism because the former defines the fourth ordered field equations, whereas the latter defines the  second-order field equations, like the conventional equations of general relativity\cite{12}. Hence, the resulting equations are not purely metric and prevail a series of algebraic relations between the fields of matter and affine connections. Many studies have been done on the mentioned formalisms, and it has been shown that in metric formalism, the scalar field appearing in the form of $\Phi=\frac{df}{dR}$ depends on the form of $f(R)$ Lagrangian while having a non-vanishing trace equation\cite{A}. This scalar field in a wide range of cosmic structures has a very low mass and its characteristics are severely constrained by cosmic and laboratory observations, at short scales\cite{13,14,15}. But, the scalar field in the version of Palatini formalism has more specific properties; for example, although it satisfies the late-time cosmic speed-up models, nevertheless, it has gradient instabilities at various theories of physics such as cosmological perturbations\cite{16,17,18}. Metric $f(R)$
and Palatini $f(\mathcal{R})$ theories may have interesting results in cosmological studies, and they may also suffer from some problems. Also, to establish a bridge between these various theories, and to avoid of possible problems, particular hybrid combinations can be considered between these two different approaches or other combinations using Lagrangian and scalar curvature.\cite{19,20,21} To get the best theory, we can use the hybrid combination of Einstein-Hilbert action with
curvature scalar $R$, and non-metric Palatini $f(\mathcal{R})$ gravity which  has been used in some cosmological studies,  focussing
on special physical and cosmological applications\cite{19,20,21,22,23}. It is clearly indicated in the literature that such combinations could easily help to explain the late-time accelerated expansion of the universe, which helps explaining dark energy and dark matter.\cite{iii} Our main goal in this article is to investigate the inflation scenario with the help of this formalism and from the Swampland conjecture point of view, which has not been studied so far. {Of course, the relationship between purely low-energy theories such as $f(R)$ gravity and swampland conjectures have also been recently discussed in literature, with interesting results. In fact, by challenging the refined swampland conjecture, the compatibility of these models with mentioned conjecture has been discussed. It has been argued that if these conjectures are correct, the effective theories live in landscape instead of swampland\cite{I}}. {{So far, various theories have been introduced to understand the different structures of this universe; among different scenarios, one of the excellent tools to create the primordial density perturbations that are comprehended as the primary candidate for the formation of universe and other issues such as the temperature anisotropy of the cosmic background photons is called Inflation\cite{bbb}.  Moreover, in the standard view, the primary responsibility to push the inflationary era is a single scalar field\cite{ccc}}}. Many inflation models have been studied to shed light on the various cosmological concepts such as inflation,  dark matter, and dark energy.\cite{aaa} Inflation periods are also divided into stages such as slow-roll and reheating, so that in the final stage of inflation the scalar field  creates particles and then the reheating phase starts. {{Among the various types of inflationary scenarios, it has recently been specified that warm inflation is well compatible with swampland criteria\cite{II,H,K}}}. The hot inflation model also explains that the radiations arise naturally without any reheating phase\cite{B}. Cosmic inflation has also been studied from different perspectives, including higher-derivative curvature theories regarding natural inflation, k-inflation, brane inflation, etc\cite{24,25,26,27,28,29,30,31,32,33}. Each inflation model can also be explained by two essential cosmological parameters, the scalar spectral index ($n_{s}$) and tensor-to-scalar ratio ($r_{s}$). Bearing all  these in mind, we want to challenge with a particular hybrid gravity theory from the Swampland conjecture perspective.\cite{34,35,36,37,38,39,40,41,42,43,44,45} To the authors knowledge, such a structure has not yet been studied for inflation models. Since  two decades ago, Swampland conjectures, weak gravity conjectures, and general Swampland programs have been seriously studied in various topics of cosmological studies such as inflation, black hole physics, dark energy and dark matter, brane structure, and other cosmic concepts\cite{44,45,46,47,48,49,50,51,52}. Recently, ones have developed various inflation models, including the modified theory of gravity such as $f(R)$, $f(R, T)$, etc., subject to various conditions as slow-roll and constant-roll, from the point of view of different Swampland conjectures\cite{C}.  In the Swampland program, we face the Swampland and the landscape, as the set of theories compatible with quantum gravity. For more information, see  Ref.s \cite{34,35,36,37,38,39,40,41,42,43,44,45,46,47,48,49,50,51,52,53,54,55,56,57,58,59}.

 We organize this paper as follows.
In section 2, we overview the field equations of hybrid metric-Paltini theory and cosmic parameters. In section 3, we introduce the inflationary parameters and investigate their cosmological implications, and compare them with cosmological observations. In this regard, we first introduce our model, and by selecting a suitable potential, we obtain cosmological quantities such as slow-roll parameters, scalar spectral index,
tensor-to-scalar ratio, etc. Then, we compare the compatibility or incompatibility of the model with observable data such as Planck by applying Swampland conjecture, {{($ C_ {1} $) and ($ C_ {2} $) being the first and second components of dS swampland conjecture}}, to the plots $r_{s}-n_{s}$, $C_{1,2}-n_{s}$ and $C_{1,2}-r_{s}$. Finally, the paper ends with a conclusion
in section 4.

\section{Overview of field equations in hybrid metric-Paltini theory}

In this section, we study the details of the hybrid gravity theory, as a hybrid combination of Einstein-Hilbert action with curvature scalar $R$ and a function $f(\mathcal{R})$ in Palatini gravity\cite{60,68',61,62,63,64,65,66}. The action of this theory is given by

\begin{equation}\label{1}
S=\frac{1}{2\mathcal{K}^{2}}\int d^{4}x\sqrt{-g}(R+f(\mathcal{R})),
\end{equation}
where $R$ and $\mathcal{R}$  are  (Einstein) scalar curvature and (Palatini) scalar curvature, respectively. The field equations  are obtained by applying the variation of action (1) with respect to the metric and connections independently, as follows\cite{60,68',61,62,63,64,65,66}:
\begin{equation}\label{2}
G_{\mu\nu}+F(\mathcal{R})\mathcal{R}_{\mu\nu}-\frac{1}{2}f(\mathcal{R})g_{\mu\nu}=0,
\end{equation}

\begin{equation}\label{3}
\nabla_{\sigma}(\sqrt{-g}F(\mathcal{R})g^{\mu\nu})=0,
\end{equation}
where $F(\mathcal{R})=\frac{df(\mathcal{R})}{d\mathcal{R}}$. Also $G_{\mu\nu}$ and $\nabla_{\mu}$ are Einstein tensor and the covariant derivative with respect to the Palatini connection $\overline{\Gamma}$, respectively \cite{60,61,62,63,66,67}. The Palatini curvature tensor is calculated using
 the Levi-Civita connection  $\overline{\Gamma}$  as {\cite{68'}}
\begin{equation}\label{4}
\mathcal{R}_{\mu\nu}=\overline{\Gamma}^{\sigma}_{\mu\nu,\sigma}-\overline{\Gamma}^{\sigma}_{\mu\sigma,\nu}+\overline{\Gamma}^{\sigma}_{\sigma\lambda}\overline{\Gamma}^{\lambda}_{\mu\nu}-\overline{\Gamma}^{\sigma}_{\mu\lambda}\overline{\Gamma}^{\lambda}_{\sigma\nu}.
\end{equation}
The Palatini scalar can be obtained by contracting as $g^{\mu\nu}\mathcal{R}_{\mu\nu}=\mathcal{R}$. By examining equation (3), we can see the compatibility of the metric tensor by a conformal factor $F(R)$.  Hence, the relation between the Ricci and Palatini curvature tensors  is expressed as follows:

\begin{equation}\label{5}
\mathcal{R}_{\mu\nu}=R_{\mu\nu}+\frac{3}{2F(\mathcal{R})}F(\mathcal{R})_{,\mu}F(\mathcal{R})_{,\nu}-\frac{1}{F(\mathcal{R})}\nabla_{\mu}F(\mathcal{R})_{,\nu}-\frac{1}{2F(\mathcal{R})}g_{\mu\nu}\Box F(\mathcal{R}),
\end{equation}

which results in the following relation between the Ricci and Palatini scalars, as

\begin{equation}\label{6}
\mathcal{R}=R+\frac{3}{2}\left[\left(\frac{\partial F(\mathcal{R})}{F(\mathcal{R})}\right)^{2}-\frac{2\Box F(\mathcal{R})}{F(\mathcal{R})}\right].
\end{equation}

Putting the Palatini curvature tensor in equation (2), results in

\begin{equation}\label{7}
\begin{split}
&(1+F(\mathcal{R}))G_{\mu\nu}=\nabla_{\mu}F(\mathcal{R})_{,\nu}-\Box F(\mathcal{R})g_{\mu\nu}-\frac{3}{2}\frac{F(\mathcal{R})_{,\mu}F(\mathcal{R})_{,\nu}}{F(\mathcal{R})}\\
&+\frac{3}{4}\frac{F(\mathcal{R})_{,\sigma}F(\mathcal{R})^{,\sigma}g_{\mu\nu}}{F(\mathcal{R})}-\frac{1}{2}(F(\mathcal{R})\mathcal{R}-f(\mathcal{R}))g_{\mu\nu}.
\end{split}
\end{equation}
Also, by contracting Eq.(2) with $g^{\mu\nu}$, one can obtain
\begin{equation}\label{8}
F(\mathcal{R})\mathcal{R}-2f(\mathcal{R})=R
\end{equation}
By  obtaining algebraically  $\mathcal{R}$ in terms of $R$ , namely $\mathcal{R}=h(R)$, the field equations for such a theory will be expressed in the following form:
\begin{equation}\label{9}
\begin{split}
(1+G(R)[( h(R))']^{-1})G_{\mu\nu}=&\nabla_{\mu}(G(R)[( h(R))']^{-1})_{,\nu}-\Box (G(R)[( h(R))']^{-1})g_{\mu\nu}\\&-\frac{3}{2}\frac{ (G(R)[( h(R))']^{-1})_{,\mu} (G(R)[( h(R))']^{-1})_{,\nu}}{ (G(R)[( h(R))']^{-1})}\\
&+\frac{3}{4}\frac{ (G(R)[( h(R))']^{-1})_{,\sigma} (G(R)[( h(R))']^{-1})^{,\sigma}g_{\mu\nu}}{ (G(R)[( h(R))']^{-1})}\\&-\frac{1}{2}(G(R)[(\ln h(R))']^{-1}-g(R))g_{\mu\nu},
\end{split}
\end{equation}
where $g(R)\equiv f(h(R)),G({R})=\frac{dg({R})}{d{R}}$ and $'$ denotes derivative
with respect to $R$. Now, we use the definitions $\Phi=G(R)[( h(R))']^{-1}$ and $V(\Phi)=G(R)[(\ln h(R))']^{-1}-g(R)$ which result in the following Einstein equations non-minimally coupled to
scalar field
\begin{equation}\label{10}
\begin{split}
(1+\Phi)G_{\mu\nu}=\nabla_{\mu}\Phi_{,\nu}-\Box\Phi g_{\mu\nu}-\frac{3}{2}\frac{\Phi_{,\mu}\Phi_{,\nu}}{\Phi}+\frac{3}{4}\frac{\Phi_{,\sigma}\Phi^{,\sigma}g_{\mu\nu}}{\Phi}
-\frac{1}{2}V(\Phi)g_{\mu\nu}.
\end{split}
\end{equation}

\subsection{Cosmological parameters in hybrid model}

Cosmic inflation can be investigated using the field equations \eqref{10}.
In this regard, we use the Friedmann-Robertson-Walker (FRW)  line element
\cite{63,64,65,66,67,68,12345,69}

\begin{equation}\label{11}
ds^{2}=dt^{2}-a(t)^{2}(dx^{2}+dy^{2}+dz^{2}),
\end{equation}
where $a(t)$ is scale factor.Using \eqref{11}, the hybrid field equations are given by
\begin{equation}\label{14}
3H^{2}(1+\Phi)+\frac{3}{4\Phi}\dot{\Phi}^{2}+3H\dot{\Phi}+V(\Phi)=0,
\end{equation}
\begin{equation}\label{13}
(1+\Phi)(3H^{2}+2\dot{H})-\frac{3}{4\Phi}\dot{\Phi}^{2}-V(\Phi)+\ddot{\Phi}+2H\ddot{\Phi}=0,
\end{equation}
where $H=\frac{\dot{a}}{a}$ is the Hubble parameter which has the dynamics
$\dot{H}+H^{2}=\frac{\ddot{a}}{a}$  .
Taking time derivative of the equation \eqref{14} and adding it up with the addition
of \eqref{14} and \eqref{13}, multiplied by $3H$, one can obtain the scalar
field equation \begin{equation}\label{16}
\ddot{\Phi}+3H\dot{\Phi}-\frac{\dot{\Phi}^{2}}{2\Phi}+\frac{\Phi}{3}(R- V'(\Phi))=0,
\end{equation}
where
 $R=6(\dot{H}+2H^{2})$ is the Ricci scalar for FLRW metric in Einstein-Hilbert
action. Now,
we rewrite $R$ in terms of the scalar field dynamics. In doing so,  one may
use  the following relation between $R$ and potential $V(\Phi)$\cite{60,61,62,63,64,65,66,67,68,12345,69}
\begin{equation}\label{18}
R=2V(\Phi)-\Phi V'(\Phi).
\end{equation}
Substituting for $R$, from  \eqref{18}, in \eqref{16}  we have
\begin{equation}\label{19}
\ddot{\Phi}+3H\dot{\Phi}-\frac{\dot{\Phi}^{2}}{2\Phi}+\frac{\Phi}{3}\Big(2V(\Phi)-(1+\Phi) V'(\Phi)\Big)=0.
\end{equation}
Applying the slow-roll regime, $V(\Phi)\gg\dot{\Phi}^{2}, 3H\dot{\Phi}\gg\ddot{\Phi}$ and $\Phi\ll 1$, one can obtain

\begin{equation}\label{20}
\dot{\Phi}=\frac{\Phi}{3H}\left(\frac{2V(\Phi)}{3}-V'(\Phi)\right),
\end{equation}
Putting $\dot{\Phi}$ from \eqref{20} in \eqref{14} and \eqref{13},  and imposing
slow-roll regime we obtain
\begin{equation}\label{22}
3H^{2}=\frac{1}{1+\Phi}\left(\frac{2V(\Phi)\Phi}{3}-(\Phi-\frac{1}{2})V'(\Phi)\right),
\end{equation}
and
\begin{equation}\label{21}
\dot{H}=3\dot{\Phi}^{2}\frac{\Big(V(\Phi)+V'(\Phi)\Big)(\Phi+\frac{3}{2})}{6\Phi V(\Phi)-4\Phi V'(\Phi)}.
\end{equation}

 Inflation can be achieved,
in this model, using the quasi de Sitter results provided that the value of the Hubble parameter is almost constant. In this regard, we can express the number of e-folds
\begin{equation}\label{23}
N=\int_{t}^{t_{e}}H dt=\int_{\Phi}^{\Phi_{e}}\frac{H}{\dot{\Phi}}d\Phi,
\end{equation}
where  $\Phi_e$ is the value of inflaton field at the end time $t_e$ of  inflation, and $H$ and $\dot{\Phi}$ are given in the aforementioned equations.
In the cosmic inflation studies, it is necessary to introduce cosmological parameters such as slow-roll, scalar spectral index, and tensor-to-scalar ratio\cite{67,68,12345,M,69}. {Slow-roll parameters are used to calculate and determine the most important cosmological quantities, such as the scalar spectrum index and the tensor-to-scalar ratio\cite{67,68,69,L}.
 Of course, there are two completely different approaches for slow-roll approximation.
The first model imposes constraints on the shape of the potential and requires that the evolution of the scalar field be asymptotic, which is used as a suitable approach to examine inflation at a particular potential\cite{67,68,69,L}.
 The second form is an approximation of the conditions applied to the Hubble parameter's evolution in the inflation period. This model has more apparent geometrical interpretations than the previous model, and in a way, its analytical features are more accessible.
For further reading,  see\cite{L}.} {{When supplied with a potential $V(\phi)$  to create an inflation model, the slow-roll approximation is typically reported as a necessity for the smallness of the two parameters represented by \cite{ttt}. Hereafter, one can refer to them as potential-slow-roll (PSR) parameters $\big[\epsilon_{1}=\alpha\big(V'/V\big)^{2}, \epsilon_{2}=2\alpha\big(V''/V\big)\big]$
where $\alpha=m_{pl}^{2}/16\pi$}}.  For this reason, we introduce  two slow roll parameters, {{$(\epsilon_{1}<0.0097 (0.0044) (95 \% CL), \epsilon_{2}<0.032^{+0.009}_{-0.008} (0.035 \pm 0.008) (68 \$ CL)$, from Planck, TT,TE,EE+lowE+lensing(BK15))}} \cite{J}
as
\begin{equation}\label{24}
\epsilon_{1}=-\frac{\dot{H}}{H^{2}},\hspace{1cm}\epsilon_{2}=\frac{\ddot{\Phi}}{H\dot{\Phi}}.
\end{equation}
Using \eqref{21} and \eqref{22}, these parameters for the hybrid gravity theory are given by,

\begin{equation}\label{25}
\epsilon_{1}=\frac{9\Big(V'(\Phi+\frac{3}{2})+V\Big)\dot{\Phi}^{2}}{\Big(4V'\Phi-6V\Phi\Big)\Big(\frac{2}{3}V'\Phi-V(\Phi-\frac{1}{2})\Big)},
\end{equation}

\begin{equation}\label{26}
\epsilon_{2}=\frac{1+\Phi}{V(\Phi-\frac{1}{2})-\frac{2V'\Phi}{3}}\left(\frac{-4\Phi V'\Big(V-\frac{2V'}{3}\Big)}{3V'\Big(\Phi+\frac{3}{2}+V\Big)}-(\Phi-\frac{2}{3})V'+V\right).
\end{equation}
In order for the inflation to work properly,
 the slow-roll  parameters should be very small. To reach these criteria,
 we have to consider  $\dot{H}<0$, which results in $0 <\epsilon_{1}\ll 1$ and $ 0<\epsilon_{2}\ll 1$. Inflation also ends when $\epsilon_{1}\approx1$ is met. {{Primordial fluctuations are density deviations in the early universe which stand assumed as the seeds of universe structures \cite{eee,fff,ggg,67,68,69,L,ttt,70,71}. Presently, the numerous widely accepted answer for their source stands in cosmic inflation systems. Concerning the inflationary scheme, the exponential expansion of the scale factor during inflation-induced inflation field quantum fluctuations is extended to macroscopic scales\cite{eee,fff,ggg,ttt,70,71}. Also, upon exiting the horizon, it freezes in. In the afterward phases of radiation-matter-domination, these fluctuations are re-entered on the horizon. Consequently, be formed the primary requirements for creating the universe configuration\cite{eee,fff,ggg,67,68,69,L,ttt,70,71}.
The primordial fluctuations' statistical characteristics can stand concluded from anisotropies observations in the CMB and distribution of matter measurements viz galaxy redshift surveys\cite{eee,fff,ggg,67,68,69,L,ttt,70,71}. Since inflation causes fluctuations, these measurements can also impose restrictions on parameters within the inflationary scenario. power spectrum that offers the variations power as a function of spatial scale quantifies the primordial fluctuations normally. In this method, one usually assumes the fractional energy density of fluctuations. For scalar fluctuations, $n_{s}$ is considered as the scalar spectral index\cite{eee,fff,ggg,67,68,69,L,ttt,70,71}. The $n_{s}$ illustrates how the density fluctuations differ with scale. Since the extent of these fluctuations relies upon the inflaton's activity when these quantum fluctuations are evolving super-horizon sized, various inflationary potentials anticipate different spectral indices. These count upon the slow-roll parameters, particularly the gradient and curvature of the potential\cite{eee,fff,ggg,67,68,69,L,ttt,70,71}.
For example, we can assume the models that curvature is positive large, so $n_{s}>1$. But, we can consider some models with unique features, such as monomial potentials indicating a red spectral index viz  $n_{s}<1$. With all of the explanations, the recent observable data such as Planck predict the $n_{s}=0.96$\cite{J,67,68,69,L,ttt,70,71}. Many inflationary models predict the existence of primordial tensor fluctuations. As mentioned earlier and with scalar fluctuations, we expect the tensor fluctuations to obey a power law and tensor index parameterize them and CMB data from the Planck satellite gives a constraint of $r<0.064$\cite{eee,fff,ggg,67,68,69,L,ttt,70,71}.}}

{{To analyze the primordial power spectrum of any inflation scenario,  three independent parameters are utilized in the literature, including the power spectrum of curvature perturbation, tensor-to-scalar ratio, and  scalar spectral index, defined as follows}},
\begin{equation}\label{27}
P_{R}=\frac{H^{2}}{8\pi^{2} \epsilon_{1}M_{pl}^{2}},\hspace{0.5cm}P_{T}=\frac{2}{M_{pl}^{2}}\left(\frac{H}{2\pi}\right)^{2},\hspace{0.5cm}r=\frac{P_{T}}{P_{R}},\hspace{0.5cm}n_{s}-1=\frac{d\ln P_{R}}{d\ln k},
\end{equation}
where $d\ln k=dN$. All of these cosmological parameters are observable so that in many studies, the values of these parameters are compared with observable data\cite{70,71}. Therefore, deep relations can be established between high-energy physics and the observational cosmology such as cosmic microwave background (CMB).  These cosmological parameters can  be obtained in the following form for the hybrid gravity theory, as
{\begin{equation}\label{29}
\begin{split}
&\mathcal{X}_{1}=3\Phi\Big(1+\Phi\Big)^{2}\Big(2\Phi+3\Big)\Big[3V(\Phi)-2V'(\Phi)\Big]\Big(V(\Phi)+V'(\Phi)\Big),\\
&\mathcal{X}_{2}=\Big(2\Phi V'(\Phi)-3\Phi V(\Phi)+\frac{3}{2}V(\Phi)\Big)^{2},\\
&r_{s}=\frac{\mathcal{X}_{1}}{\mathcal{X}_{1}},
\end{split}
\end{equation}
\begin{equation}\label{28}
\begin{split}
&\mathcal{Y}_{1}=\Big(2\Phi V'(\Phi)-3\Phi V(\Phi)+\frac{3}{2}V(\Phi)\Big)^{3},\\
&\mathcal{Y}_{2}=\Phi\Big(1+\Phi\Big)^{3}\Big(2\Phi+3\Big)\Big[3V(\Phi)-2V'(\Phi)\Big]\Big(V(\Phi)+V'(\Phi)\Big),\\
&P_{R}=-\frac{1}{54\pi^{2}M_{pl}^{2}}\left(\frac{\mathcal{Y}_{1}}{\mathcal{Y}_{2}}\right),
\end{split}
\end{equation}}
{\begin{equation}\label{30}
\begin{split}
n_{s}-1=&\frac{3\Phi(1+\Phi)^{\frac{3}{2}}\Big(2V'(\Phi)-3V(\Phi)\Big)}{\Big(2\Phi V'(\Phi)-3\Phi V(\Phi)+\frac{3}{2}V(\Phi)\Big)^{\frac{3}{2}}}\bigg\{3\bigg(2\Phi V''(\Phi)-3V(\Phi)-3\Phi V'(\Phi)  \\
&+\frac{7}{2}V'(\Phi)\bigg)\bigg[2\Phi V'(\Phi)-3\Phi V(\Phi)+\frac{3}{2}V(\Phi)\bigg]^{-1}-\frac{3}{1+\Phi}-\frac{4\Phi+3}{\Phi(2\Phi+3)}\\
&-\frac{3V'(\Phi)-2V''(\Phi)}{3V(\Phi)-2V'(\Phi)}-\frac{V''(\Phi)+V'(\Phi)}{V'(\Phi)+V(\Phi)}\bigg\}.
\end{split}
\end{equation}}

By adding a tensor component such as a single-parameter extension of $\Lambda$CDM, the tensor-to-scalar ratio is found {{$r_{0.002}<0.10 (95 \% CL, Planck TT+lowE+lensing)$\cite{J,70,71}, Also $r_{0.002}<0.056 (95 \% CL, Planck TT,TE,EE+lowE+lensing+BK15)$\cite{J}, and $r_{0.002}<0.064 (95 \% CL, Planck TT,TE,EE+lowE+lensing+BK15 and Virgo2016)$\cite{J}}} which is
in agreement with the recent observations.  Of course, if we consider the analysis of Planck 2013 observable data\cite{70,71}, the Hubble constant value is {{$67.36 \pm 0.54 km/sMpc^{-1}$ from  TT,TE,EE+lowE+lensing}}\cite{J}. In simple inflation models,  the running of  scalar spectral index has  second-order dependence on slow-roll parameters and is very small {{($\frac{dn_{s}}{d\ln k})\approx-0.0045 \pm 0.0067 (68 \% CL)$ from Planck TT,TE,EE+lowE+lensing\cite{J}}}.  {Also, we can check these important parameters according to the latest Planck data, which is based on an error of 68 percent. On the other hand, we give the $n_{s}$ and $r$ constraints arising from the marginalized joint 68\% and 95\% CL regions of the Planck 2018 in combination with BK14+BAO data,i.e., $n_{s}=0.9649\pm0.0042$ at 68\% Cl, and $r<0.1$ at 95\% CL and from BICEP2/Keck Array BK14 recent data $r<0.056$ at 95\% CL \cite{J}.} In the next section, we investigate
the inflation model and its cosmological parameters in the framework of hybrid gravity theory accompanied
by Swampland conjectures,
   to  examine its  compatibility  with  observable data.

\section{Inflation in hybrid metric-Paltini theory and \\Swampland conjectures}
{{A maximally symmetric Lorentzian manifold with constant positive scalar curvature can introduce a de Sitter space summarized as dS\cite{lll,nnn,mmm,ooo} Actually, it is an $n$-sphere Lorentzian analogue, including its canonical Riemannian metric. The primary application of dS is its usefulness in general relativity\cite{lll,nnn,mmm,ooo}. It acts as one of the most straightforward mathematical cases of the universe compatible with the observed accelerating expansion of the universe. Mainly, dS space is the maximally symmetric vacuum solution of Einstein's field equations with a positive cosmological constant\cite{lll,nnn,mmm,ooo}. There exists  cosmological proof that the universe itself is asymptotically de Sitter, i.e., it will evolve like the de Sitter universe in the far future when dark energy dominates\cite{lll,nnn,mmm,ooo}.}}
{{In this section, we want to challenge the inflation model in this particular hybrid gravity theory from the Swampland conjecture perspective. We investigate its cosmological implications and compare them with cosmological observations. Therefore, we first introduce our structure, and by selecting a suitable potential, we obtain cosmological quantities such as slow-roll parameters, scalar spectral index, tensor-to-scalar ratio, etc. Then, we compare the compatibility or incompatibility of it with the latest observable data, such as Planck\cite{J}, by applying Swampland conjectures to some plots, $r_{s}-n_{s}$ plan, $C_{1,2}-n_{s}$ and $C_{1,2}-r_{s}$. Recently, Swampland Conjectures have been particularly important in effective low-energy physical hypotheses\cite{34,35,36,37,38,39,40,41,42,43,44,45,46,47,48,49,50,51,52,53,54,55,56,57,58,59}. Their applications to inflationary cosmology, black hole physics, and even dark matter and dark energy have been widely studied\cite{50,51,52,53,54,55,56,57,58,59}. In the literature, there are different gravitational samples along with the swampland program, and a series of them, such as the slow-roll single-field inflation mechanism, show inconsistencies with the second condition of the dS Swampland Conjecture\cite{34,35,36,37,38,39,40,41,42,43,44,45,46,47,48,49,50,51,52,53,54,55,56,57,58,59}. However, several approaches have led to compatibility with Swampland Conjecture by modifying the original systems and studying the inflation on Brane and from the Gauss-Bonnet parameter monitoring\cite{K, D, 48}.
When the original sample is somewhat modified, it shows consistency with the Swampland Conjectures, an example of which is a recent study of cosmological parameters in the logarithmic model using Swampland Conjecture. In this regard, we intend to study whether a modified gravity like Hybrid metric-Palatini gravity can show consistency with the Swampland Conjectures. The general goal of the authors is to find a particular class of ideas or a specific classification of approaches in line with the swampland program derived from string theory, which can be proved to confirm the string theory \cite{50,51,52,53,54,55,56,57,58,59}. However, in this article, we apply the refined Swampland Conjecture to the mentioned model, expressed in the following form}}\cite{D,40,41,42,43,45,44,46,47,48,49,50}

\begin{equation}\label{31}
M_{pl}\frac{|V'|}{V}>C_{1},\hspace{1cm}M_{pl}^{2}\frac{|V''|}{V}<-C_{2},
\end{equation}
Where $C_{1}$ and $C_{2}$ are constant parameters of order 1,i.e., $\mathcal{O}(1)$. {{Here, we want to analyze the cosmological perturbations near homogeneous FLRW space-time. Our universe structures have been created from small seed perturbations, then they were developed, evolved, and became visible. The quantum fluctuation is the most suitable surmise of the origin of these perturbations during inflation\cite{E}. This inflationary epoch creates scopes of an extremely short scale to the size of the visible universe conducting to the gravitational waves and density perturbation\cite{F}. Inflation fields generate small fluctuations called quantum fluctuations. They are related to the space-time metric fluctuations offering advancement to the curvature $R$ perturbations. These curvature perturbations can be assumed as the gravitational potential. Also, they can be related to the perturbation of matter via field equations of our approaches\cite{G}. According to all the mentioned points, consequently, the perturb of the effective potential is as follows,}}

\begin{equation}\label{32}
V(\Phi)=V_{0}+V_{1}\Phi^{n}.
\end{equation}
{The constant parameters used in the
potential are noteworthy about the physical concepts. For the case $(n= 2)$, the trace equation  automatically implies  $R=-2\mathcal{K}T+2V_{0}$\cite{M,12345}
which in the limit $T\rightarrow0$, accompanied by the cosmic expansion, is naturally evolved to the dS phase where  $V_{0}\sim\Lambda$ is required for consistency with the latest observations.
Also, if  $V_{1}>0$, the dS regime represents the minimum of the potential.
Then, the effective mass required for the local experiments, namely $m^2\Phi=2(V_{0}-2V_{1}\phi)/3$
will be positive and small, provided that  the inequality $\phi<V_{0}/V_{1}$ is satisfied. For huge $V_{1}$, the field amplitude can be reduced to fit the Solar System tests.
 Of course, the points made by the authors in their papers \cite{12345,M}  were that the exact dS solutions be compatible with the scalar field dynamics in this study.
Ones on galactic dynamics believe that the hybrid metric-Palatini model could be a viable alternative to the dark matter paradigm of cosmology today \cite{M,12345}.
Now, by mentioning these explanations and concepts, by selecting the power of order $n$, which is a generalization of the mentioned model, and combining it with swampland conjectures, we should somehow seek to choose the most appropriate range for the compatibility of all the above concepts.
 Regarding selecting the constant parameter $(n<2)$, the chosen values had unacceptable results for the cosmological parameters of the scalar spectrum index and the tensor-to-scalar ratio.
 Therefore, by selecting any of these cosmological parameters, including the ones mentioned above, the range of the experiment and the desired amplitude of the scalar field will change. In addition to the mentioned cases,  see\cite{68'}
for further study.}

Using (28) and (29), one can obtain
\begin{equation}\label{33}
\frac{n\Phi^{-1+n}V_{1}}{V_{0}+\Phi^{n}V_{1}}>C_{1},
\end{equation}

\begin{equation}\label{34}
\frac{(-1+n)n\Phi^{-2+n}V_{1}}{V_{0}+\Phi^{n}V_{1}}<-C_{2}.
\end{equation}

According to the first field equations, the Hubble parameter is calculated as follows
\begin{equation}\label{35}
H=\sqrt{\frac{\frac{2}{3}nV_{1}\Phi^{n}-(-\frac{1}{2}+\Phi)(V_{0}+\Phi^{n}V_{1})}{3(1+\Phi)}}.
\end{equation}

The slow-roll parameters introduced in equations (22) and (23)  are obtained using this form of  potential function
as\begin{equation}\label{36}
\epsilon_{1}=-\frac{3(1+\Phi)\Big(3\Phi V_{0}+\Phi^{n}(-2n+3\Phi)V_{1}\Big)\Big(2\Phi V_{0}+\Phi^{n}\Big[3n+2(1+n)\Phi\Big]V_{1}\Big)}{\Phi\Big((-3+6\Phi)V_{0}+\Phi^{n}\Big[-3-4n+6\Phi\Big]V_{1}\Big)^{2}},
\end{equation}

\begin{equation}\label{37}
\epsilon_{2}=\frac{6(1+\Phi)\bigg(V_{0}-n(-\frac{3}{2}+\Phi)\Phi^{-1+n}V_{1}+\Phi^{n}V_{1}+\frac{8nV_{1}\Phi^{n}\Big[-3\Phi V_{0}+\Big(2n-3\Phi\Big)\Phi^{n}V_{1}\Big]}{18\Phi V_{0}+9\Phi^{n}\Big(3n+2(1+n)\Phi\Big)V_{1}}\bigg)}{(-3+6\Phi)V_{0}+\Phi^{n}\Big(-3-4n+6\Phi\Big)V_{1}}.
\end{equation}

Other cosmological parameters such as the scalar spectral index ($n_{s}$) and the tensor-to-scalar ($r_{s}$) ratio can also be obtained by defining these functions in equations (25) and (27), which are expressed in the following form
\begin{equation}\label{38}
r_{s}=\frac{6(1+\Phi)^{2}(3+2\Phi)\Big(\Phi V_{0}+\Phi^{n}(n+\Phi)V_{1}\Big)\Big[3\Phi V_{0}+\Phi^{n}\Big(-2n+3\Phi\Big)V_{1}\Big]}{\Big(3\Phi(-1+2\Phi)V_{0}+\Phi^{n}\Big[-2n+3\Phi(-1+2\Phi)\Big]V_{1}\Big)^{2}},
\end{equation}

\begin{equation}\label{39}
\begin{split}
&A=6\sqrt{2}\Phi(1+\phi)^{\frac{3}{2}}\Big(-3\Phi V_{0}+(2n-3\Phi)\Phi^{n}V_{1}\Big)\bigg\{-\frac{1}{\Phi}-\frac{3}{1+\Phi}-\frac{2}{3+2\Phi}+\frac{n\Big(2-2n+3\Phi\Big)V_{1}}{-3\Phi^{2-n}V_{0}+\Big(2n-3\Phi\Big)\Phi V_{1}}\\
&-\frac{n\Big(-1+n+\Phi\Big)V_{1}}{\Phi^{2-n}V_{0}+\Phi(n+\Phi)V_{1}}+\frac{18\Phi V_{0}+3\Phi^{n}\Big(-n(3+4n)+6(1+n)\Phi\Big)V_{1}}{\Phi\Big[(-3+6\Phi)V_{0}+\Phi^{n}\Big(-3-4n+6\Phi\Big)V_{1}\Big]}\bigg\},\\
&B=\Big\{-\Phi\Big[(-3+6\Phi)V_{0}+\Phi^{n}\Big(-3-4n+6\Phi\Big)V_{1}\Big]\Big\}^{\frac{3}{2}},\\
&n_{s}-1=\frac{A}{B}.
\end{split}
\end{equation}

Other cosmic parameters introduced in the article, such as $N$, representing the number of e-folds, or power spectrum of curvature perturbations, i.e., $P_{R}$ can also be calculated according to the definitions mentioned above. But, our goal in this article is to challenge this form of gravity accompanied
by Swampland conjectures\cite{54,55,56,57,58,59}. Therefore, according to the above computational values, we will examine a number of these limitations. For this reason, we first perform a series of calculations and simple manipulations. That is, we first consider the two cosmological parameters ($n_{s}$) and ($r_{s}$), which are in terms of the $\Phi$.
We inverse these equations and create structures such as $\Phi-n_{s}$ and $\Phi-r_{s}$. Then, by placing these values in equations (30) and (31), 4 new structures are created, which are expressed in the form $C_{1,2}-n_{s}$ and $C_{1,2}-r_{s}$. These structures have very long values, so we use some figures to show them, together with the acceptable range of cosmological parameters and Swampland coefficients. Let us now examine the variations of these functions relative to each other by plotting some figures. The best fits for free parameters are $(n=3, V_{0}=0.9, V_{1}=2.1)$. These values give us the best compatibility with observational data. We did not have acceptable range values for the  $n<0$ and ($0<n<2$), but the ($2.1<n<3.9$) answers were comparable to the observable data. The case  $n=3$ has the best values for a  consistency with the observational data, which we also used to plot the figures.

\begin{figure}[h!]
\begin{center}
\subfigure[]{
\includegraphics[height=6cm,width=6cm]{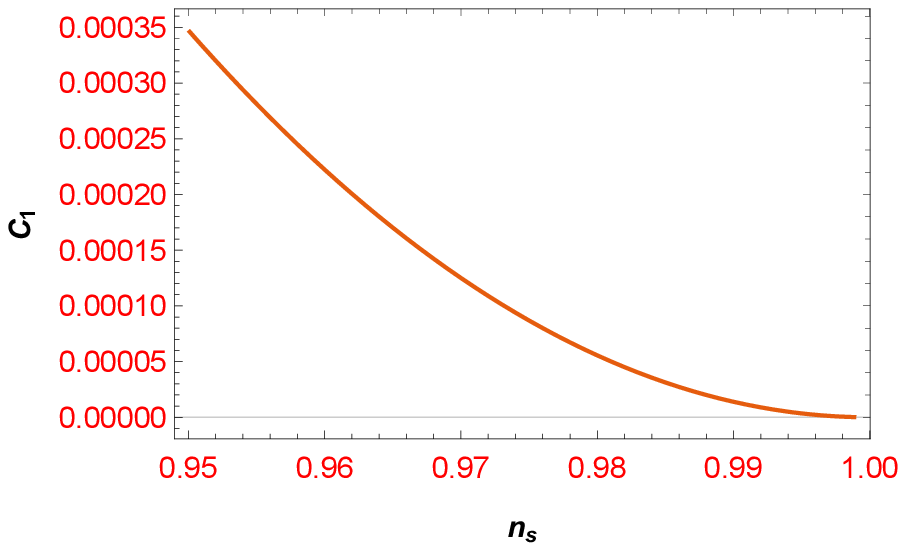}
\label{1a}}
\subfigure[]{
\includegraphics[height=6cm,width=6cm]{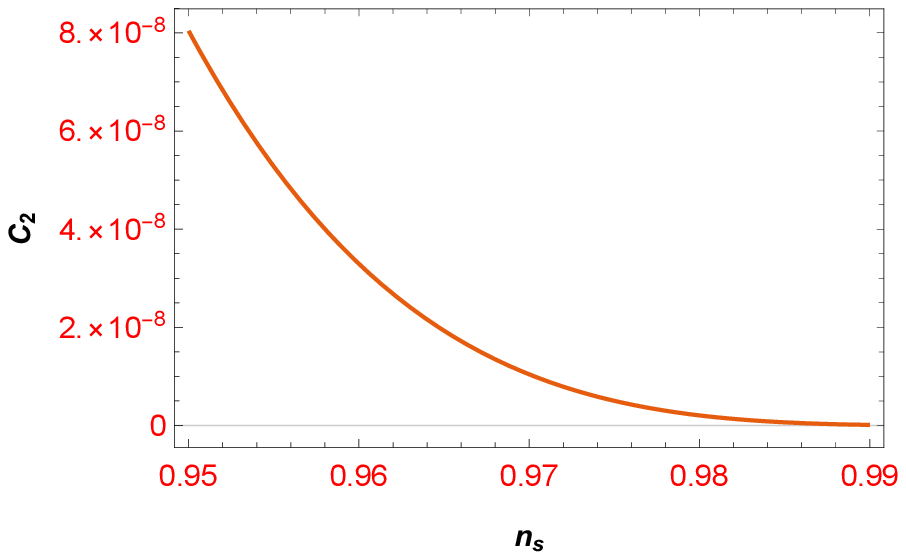}
\label{1b}}
\caption{\small{The plot of $C_{1}$ and $C_{2}$ in terms of $n_{s}$ with respect to constant parameter values as $n=3$, $V_{0}=0.9$ and $V_{1}=2.1$ }}
\label{1}
\end{center}
\end{figure}

\begin{figure}[h!]
\begin{center}
\subfigure[]{
\includegraphics[height=6cm,width=6cm]{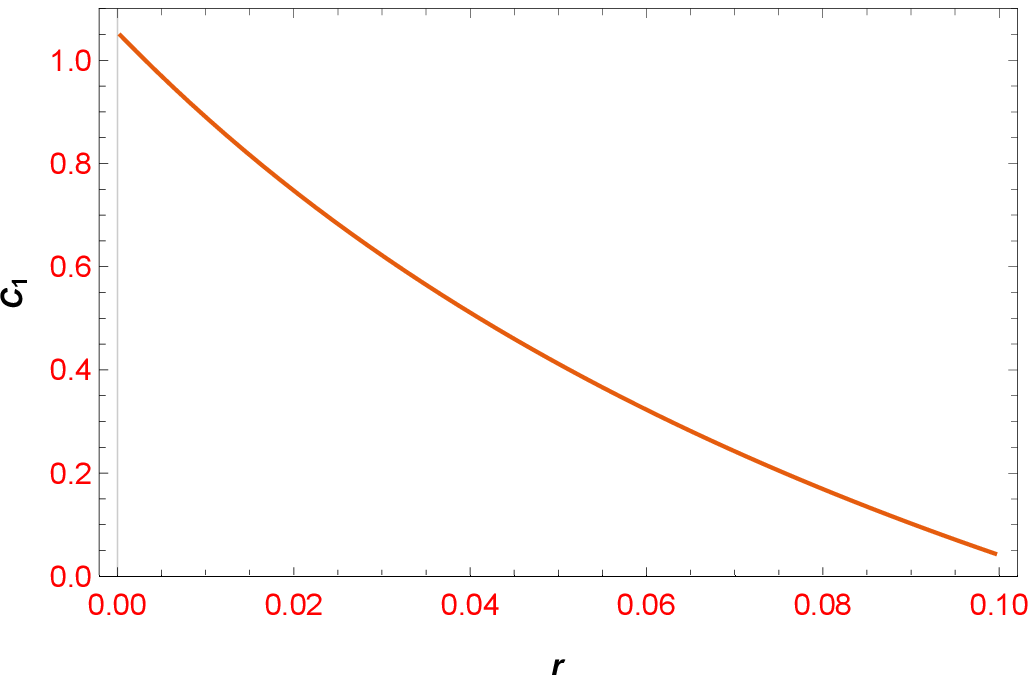}
\label{2a}}
\subfigure[]{
\includegraphics[height=6cm,width=6cm]{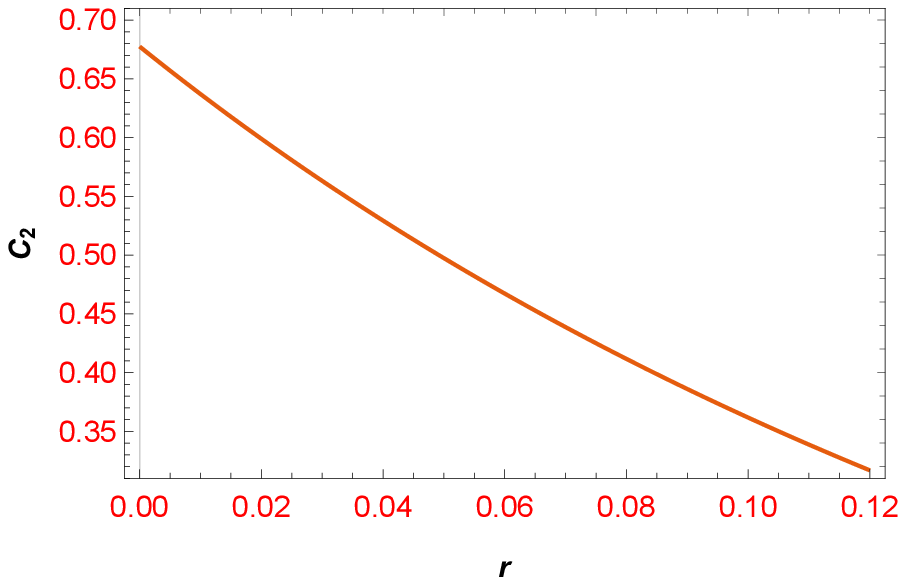}
\label{2b}}
\caption{\small{The plot of $C_{1}$ and $C_{2}$ in terms of $r_{s}$ with respect to constant parameter values as $n=3$, $V_{0}=0.9$ and $V_{1}=2.1$ }}
\label{2}
\end{center}
\end{figure}

Regarding the explanations given above, each of the following figures can be analyzed like the figure (1) to show the restrictions imposed by the Swampland conjecture, according to the cosmological parameter of the scalar spectrum index for the allowable values concerning constant parameters. As shown in the figure (1), the validate range is acceptable for both the Swampland conjecture coefficients $ C_{1} $, $ C_{2} $ and the cosmological parameter $n_{s}$. Valid values for $C_{1}$ and $C_{2}$ are shown in the figures. $ C_{2} $ must be less than $ C_{1} $\cite{K,48,I,53,54,55,57,H,AR,K}. The allowable values for these parameters are different in various theories in literature \cite{K,48,I,H,AR,K}. We can also see the same results about another cosmological parameter, i.e., the tensor-to-scalar ratio. The limitations and values allowed for this parameter, as well as the constant coefficient of the Swampland conjectures  shown in the figure (2),  indicate the compatibility of this model with a Swampland conjecture. In the following, we examine another constraint related to two cosmological parameters, namely the tensor-to-scalar ratio and the scalar spectral index, subject to the hybrid gravity and Swampland conjecture, which are well defined in the below figures.

\begin{figure}[h!]
\begin{center}
\subfigure[]{
\includegraphics[height=6cm,width=6cm]{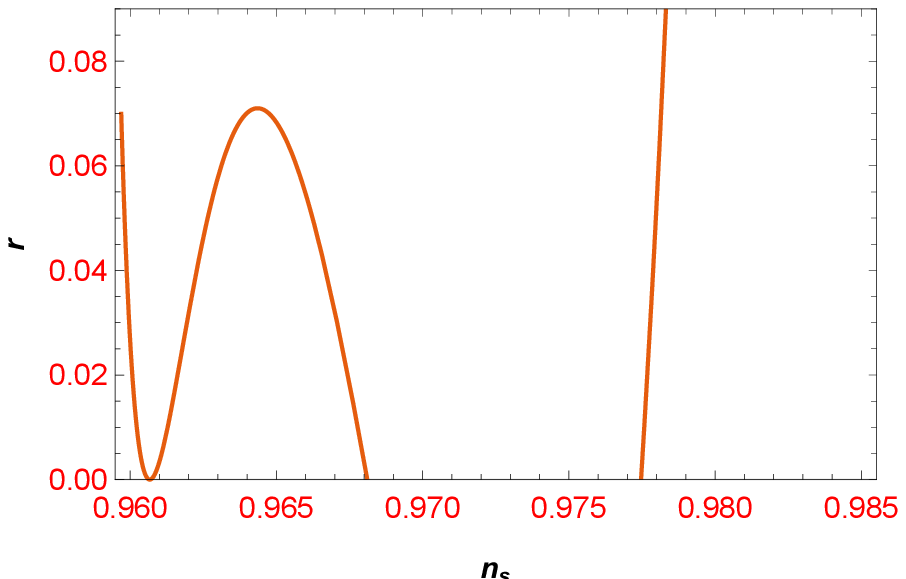}
\label{3a}}
\subfigure[]{
\includegraphics[height=6cm,width=6cm]{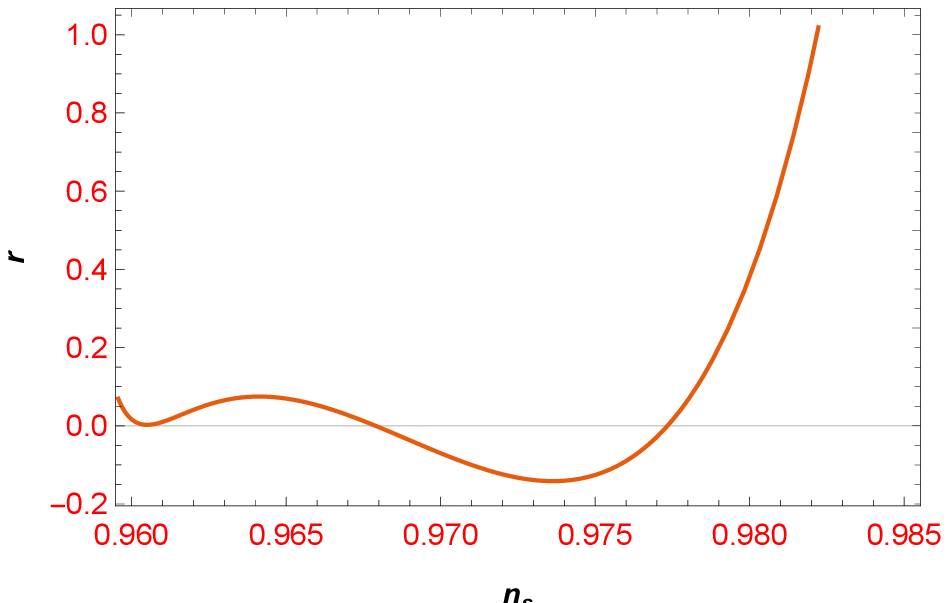}
\label{3b}}
\caption{\small{The plot of $r_{s}$ in term of $n_{s}$ with respect to constant parameter values as $n=3$, $V_{0}=0.9$ and $V_{1}=2.1$ }}
\label{3}
\end{center}
\end{figure}

{We know that the allowed values, according the latest observable data, are $n_{s}=0.9649$ and $r<0.056$\cite{70,71,J} together with the two components ($C_{1}$) and ($C_{2}$) of the order 1, i.e., $\mathcal{O}(1)$\cite{K,48,I,53,54,55,57,H,AR,K}.
Now, with a brief look, we can specify the values of each of these component ($C_{1}$) and ($C_{2}$) at acceptable points for each of these observable, i.e., $n_{s}=0.9649$ and $r<0.056$\cite{J}.
The changes to the curvilinear form of the figure are due to the continuous range of the scalar spectrum index and the tensor to scalar ratio, i.e., $ 0.95<n_{s}<0.99$ and $0<r<0.1$ concerning $1\sigma$ and $2\sigma$.
 By plotting the curves of the components ($C_{1}$) and ($C_{2}$) in terms of $n_{s}$ and $r$, we are looking for the acceptable range of ($C_{1}$) and ($C_{2}$)  concerning the mentioned constant parameters
and the most recent observational data.
To examine each of the swampland components in terms of $n_{s}$ and $r_{s}$, we rewrote the scalar field separately in terms of these two parameters according to the free parameters mentioned in the text.
The two swampland component conditions presented in the equation (28) represent a particular property, namely that they represent two different forms, viz the formulas for the refined swampland conjecture are strange. There is a kind of separation between these two conditions so that these equations can not provide any information simultaneously according to their definitions. So,  Andriot and  Roupec extended the above conjectures and introduced a new relation called further refining de Sitter swampland conjecture \cite{AR}, which can give us a complete explanation, and combine the two previous equations. Hence, this discrepancy in two different versions can be acceptable to some extent. One condition in a model may show more accurate results than another condition. In connection with
these points, as well as another issue that indicates smaller values of component $C_{2}$ than component $C_{1}$,  see the  Ref.s \cite{K,48,I,53,54,55,57,H,AR,K}.
We are investigating  the compatibility of different inflation
models with swampland conjectures, such as de Sitter swampland, according to the latest observable data\cite{48,53,54,55,56,57,H,J,K}. By studying different models,  considering the proposed conjectures, a new classification of inflation models compatible with the swampland conjectures  can be found. In case of non
applicability  of these conjectures,  we can consider updating of these conjectures or looking for new conjectures as well as correcting the desired conjectures.
 As we mentioned, such a correction has happened several times, one of which is the same as in the text.
With all these descriptions, the expression of compatibility and incompatibility in its two examples showed an important point that leads to correcting these conjectures and the presentation of stronger conjectures that reduce the rate of this incompatibility.
 Of course, there was also a discrepancy between the second condition of the swampland conjecture and its slow-roll inflation, which was tried to be improved with solutions.
The results expressed in this article can also be compared with\cite{53,54,55,57,H,AR,K,48,I}.} {{
As stated in the literature and also in this article, each of these components is constant parameters of order 1\cite{K,48,I,53,54,55,57,H,AR,K},
But, as we explained in the text, this compatibility does not exist in all points or the whole scope of the mentioned theory. In some way, there is no compatibility at all for specific values of the parameter $n<0$ and $0<n<2$.
 Hence, the dS swampland conjecture or, more precisely, any of their components have a smaller acceptable range. In other words, they have a less acceptable parametric range than warm inflation or other theories studied in the literature\cite{II,H,K}. The small parametric acceptable range of swampland conjectures can vary greatly depending on the theory. The fact that each approach has different satisfactory ranges about swampland conjectures, also more or less compatibility with the latest observable data\cite{J}, can be used to study  such theories within the acceptable range of these conjectures  consistent  with quantum gravity.
However, with the further advancement of science and access to more robust data, much more accurate results can be imagined for this idea.
But what we encounter most often in the literature right now are coefficients with constant values of unit order\cite{K,48,I,53,54,55,57,H,AR,K} that, depending on different theories, can increase or decrease the acceptable parametric range.
However, what is more important is the compatibility of all these components with observable data, including both the swampland components and all the important cosmological parameters and the specific limitations that exist for each theory in a separate form.
In fact, combining all these factors gives us the best results in our decisions.
 Of course, as we mentioned, recently, corrections have been added to this conjecture, which leads to more compatibility of these conjectures or the same components mentioned.
 However, this compatibility is in a specific parametric range, as we have already mentioned. This point can be very important and help to classify different types of theories compatible with this nascent idea}}. As mentioned above, the figure (3) shows the constitutions of two cosmological parameters. The allowable values for each of these parameters are determined according to acceptable values of the constant parameter. The interesting point is that this model, equipped by the Swampland conjectures, provides us
with good results in examining the limitations according to the observable data. {The studies and results of this paper (challenging swampland conjectures and Hybrid metric-Palatini gravity of theory) can be compared with other theories having different conditions and frameworks.
One can select the best and most accurate model by comparing the results and matching with the latest observable data.}
\section{Conclusion}
Cosmological implications of different gravitational models such as $f(R)$, $f(R, T), f(Q)$,  teleparallel gravity, and  Einstein Gauss bonnet gravity in the study of  structure of the universe with respect to different conditions and conjectures (slow-roll, constant-roll,  etc.) in various theories such as inflation, black hole physics, dark energy and dark matter have been examined and compared with observable data\cite{H,M,K,I,48}.  In this paper, we have studied a special type of gravitational theory, i.e., hybrid gravity theory (it is a hybrid combination of the Einstein-Hilbert action with curvature $R$, and a function $f(\mathcal{R})$ in Palatini gravity) in inflation from the Swampland conjecture point of view. This hybrid model has been examined in cosmological studies, and its applications have been expressed in the literature\cite{19,20,21,22,23}. Ones have used these theories as an alternative to the theories of
dark matter and dark energy\cite{19,20,21,22,23,iii}. Cosmological applications, especially inflation, in the structure of general relativity and modified effective $f(R)$ theories have been studied separately\cite{I}. Here, we have reviewed a combination of these two theories. We examined cosmological parameters such as slow-roll, scalar spectral index, and tensor-to-scalar ratio in a hybrid gravity
inflation model and challenged this particular gravity theory from the Swampland conjecture perspective\cite{K,48,I,53,54,55,57,H,AR,K}. We investigated its cosmological implications and compared them with cosmological observations. In this regard, we first introduced our model, and by selecting a suitable potential, we obtained cosmological quantities such as slow-roll parameters, scalar spectral index,
tensor-to-scalar ratio, etc. Then, we examined the compatibility or incompatibility of the model with observable data such as Planck\cite{J}, by  applying Swampland conjecture  to $r_{s}-n_{s}$ , $C_{1,2}-n_{s}$ and $C_{1,2}-r_{s}$ plots. The best fits for free parameters are $(n=3, V_{0}=0.9, V_{1}=2.1)$. These values give us the best compatibility with observational data\cite{J}. We  have no acceptable range values for   $n<0$ and ($0<n<2$), but the range ($2.1<n<3.9$)  are comparable to the observable data, and  $n=3$ gives   the best values in consistency with the observational data. All the studies in this article can be repeated for different types of modified gravitational theories and their results may be compared with the computational values obtained in this paper, as well
as the observable data. In this way, the most accurate model can be selected which is most compatible with the observable data. In the future work, we will examine other types of modified gravity models in this framework.

\end{document}